\documentclass[a4paper,superscriptaddress,notitlepage,12pt,tightenlines]{revtex4-2}
\usepackage{graphicx}
\usepackage{amsmath,amssymb,setspace}
\usepackage{newtxtext}
\usepackage{newtxmath}
\usepackage{xfrac}
\usepackage[utf8]{inputenc}
\usepackage[x11names]{xcolor}
\usepackage[unicode,colorlinks,citecolor=Blue3,linkcolor=Blue3,bookmarks=true]{hyperref}
\newcommand{\eg}{{\it e.g.\,}}
\newcommand{\hence}{\ \Rightarrow\ }
\newcommand{\partd}[2]{\dfrac{\partial#1}{\partial#2}}
\newcommand{\mean}[1]{\langle#1\rangle}
\newcommand{\smatrix}[4]{\begin{pmatrix}#1 & #2 \\ #3 & #4\end{pmatrix}}

\begin{document}

\title{Broadening the high sensitivity range of squeezing-assisted interferometers by means of two-channel detection}

\author{Gaurav~Shukla}
\affiliation{Department of Physics, Institute of Science, Banaras Hindu University,  Varanasi-221005, India}
\author{Dariya~Salykina}
\affiliation{Faculty of Physics, M.V. Lomonosov Moscow State University, 119991 Moscow, Russia}
\affiliation{Russian Quantum Center, Bolshoy Bulvar 30/bld.~1, 121205 Skolkovo, Moscow, Russia.}
\author{Gaetano~Frascella}
\affiliation{Max Planck Institute for the Science of Light, Staudtstr.~2, 91058 Erlangen, Germany.}
\affiliation{University of Erlangen-Nuremberg, Staudtstr. 7/B2, 91058 Erlangen, Germany.}
\author{Devendra~Kumar~Mishra}
\affiliation{Department of Physics, Institute of Science, Banaras Hindu University, Varanasi-221005, India}
\author{Maria~V.~Chekhova}
\affiliation{Max Planck Institute for the Science of Light, Staudtstr.~2, 91058 Erlangen, Germany.}
\affiliation{University of Erlangen-Nuremberg, Staudtstr. 7/B2, 91058 Erlangen, Germany.}
\author{Farid~Ya.~Khalili}
\affiliation{Russian Quantum Center, Bolshoy Bulvar 30/bld.~1, 121205 Skolkovo, Moscow, Russia.}
\affiliation{NUST ``MISiS'', Leninskiy Prospekt~4, 119049 Moscow, Russia.}

\begin{abstract}
For a squeezing-enhanced SU(2) interferometer, we theoretically investigate the possibility to broaden the phase range of sub-shot-noise sensitivity. We show that this goal can be achieved by implementing detection in both output ports, with the optimal combination of the detectors outputs, leading to a phase sensitivity independent of the interferometer operation point. Provided that each detector is preceded by a phase-sensitive amplifier, this sensitivity could be also tolerant to the detection loss.
\end{abstract}

\maketitle

\section{Introduction}

Optical interferometry has been an efficient tool in a plethora of experiments, starting from the well-known Michelson-Morley's refutation of the ether theory \cite{Michelson1887} to the high-resolution spectroscopy \cite{Anderson_ch35_2019} and to the gravitational-wave detection \cite{PRL_116_131103_2016}. The phase sensitivity of the state-of-art optical interferometers is limited by the quantum fluctuations of the probing light phase. In the simplest case of a coherent quantum state, the corresponding limit is known as the shot-noise one (SNL):
\begin{equation}\label{dphi_SNL}
  \Delta\phi_{\rm SNL} = \frac{1}{\sqrt{N}} \,,
\end{equation}
where $N$ is the number of quanta used for the measurement (the numerical factor could vary depending on the phase normalization).

Better sensitivity could be achieved using more sophisticated quantum states of light, see \eg the review \cite{Demkowicz_PIO_60_345_2015} and the references therein. The first practical scheme of the sub-SNL interferometer, which uses the Gaussian squeezed states,  created by a degenerate optical parametric amplifier (DOPA), was proposed by Caves in the  paper \cite{Caves1981}. Now this idea is implemented in the modern laser gravitational-wave detectors GEO-600 \cite{Nature_2011} and Advanced LIGO \cite{Nature_2013, Tse_PRL_123_231107_2019}. However, the sensitivity gain provided by the squeezed light is limited by the optical losses, both internal (inside the interferometer) and external (absorption in the output path and the detectors inefficiency). For example, due to this reason, in the Advanced LIGO, a $7.2$\,dB squeezing gave a phase sensitivity enhancement of only $3.2$\,dB \cite{Tse_PRL_123_231107_2019}.

A remedy against the detrimental effect of loss has been proposed by Caves in the same paper \cite{Caves1981}. He showed that while the influence of internal loss in the interferometer is inevitable, the external loss, including the detection inefficiency, can be overcome by amplifying the output signal by a second DOPA before detection. This phase-sensitive amplifier enhances only the quadrature carrying the phase information and thus makes it robust to loss, without introducing additional noise. Therefore, the sensitivity of the interferometer is improved.

It was assumed in the work \cite{Caves1981} that the squeeze factors of the input and the output DOPAs $r_{1,2}$ have to have equal absolute values, $|r_1|=|r_2|$ (and, evidently, have opposite signs, $r_1r_2<0$). Further developing this idea, some of us showed theoretically in Ref.\,\cite{17a1MaKhCh} and demonstrated experimentally in \cite{20a1FrAgKhCh} that the unbalanced case $|r_2|>|r_1|$ provides a better sensitivity. By sufficiently increasing the squeeze factor of the second DOPA one can, in principle, completely overcome the effect of the external loss.

Typically in high-power high-precision interferometers, \eg the gravitational-wave detectors, one of the interferometer output ports should be tuned to the dark fringe in order to eliminate the technical noise of the laser and to avoid exposing the photodetector to a high optical power. However, this regime limits the range of phases where the high sensitivity is provided, while the optical power limitation could be not so significant in smaller-scale (table-top) interferometers.

In Refs.\,\cite{Demkowicz_PIO_60_345_2015, Gard_EPJQT_4_4_2017, Ataman_PRA_98_043856_2018, Anderson_ch35_2019}, a differential detection, with the output signal proportional to the difference of photocurrents of two photodetectors located in both outputs of the interferometer, was considered. However, as we show here, this scheme also has a limited high sensitivity range.

Here, we consider the most universal approach of combining the two photodetectors outputs with the optimal weight factors. We assume that both photodetectors could be preceded by the DOPAs in order to suppress the influence of the photodetectors inefficiency and other external losses. We show that unlike the dark port regime, this method enables beating the SNL by the same amount regardless of the phase shift in the interferometer (operating point). At the same time, similar to the dark port regime, it is tolerant to the laser technical noise.

The paper is organized as follows. In Sec.\,\ref{sec:ifo} we derive the input/output equations for this interferometer and find the optimal squeeze angles for all three parametric amplifiers. In Sec.\,\ref{sec:optimization} we calculate the optimized sensitivity of this scheme and compare it with the single-detector and differential-detection cases. In Sec.\,\ref{sec:conclusion} we summarize the obtained results.

\section{Input/output relations}\label{sec:ifo}

\begin{figure}
  \includegraphics[width=0.7\textwidth]{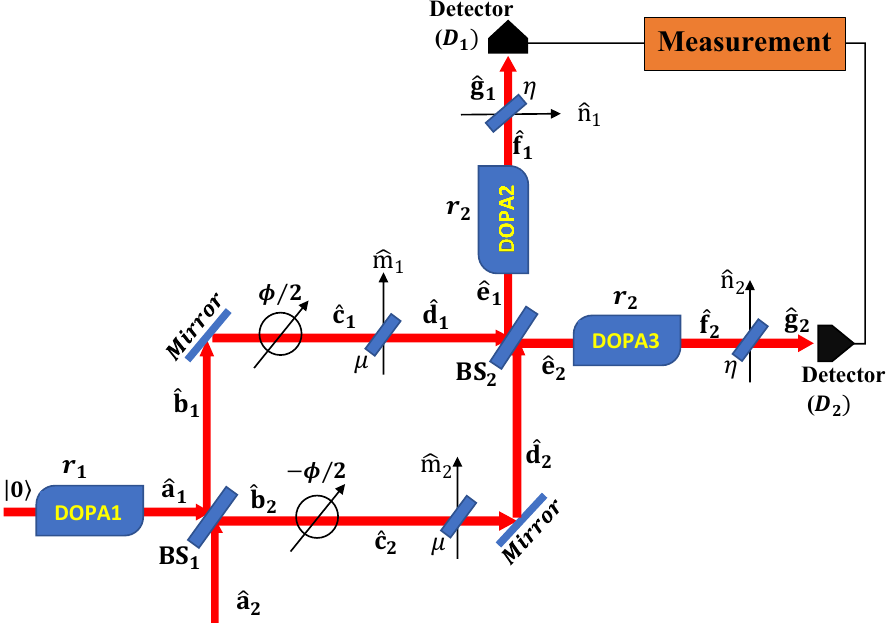}
  \caption{A Mach-Zehnder interferometer with the input degenerate optical parametric amplifiers DOPA1 (the squeezer) and two output ones DOPA2 and DOPA3. The optical losses are represented by the imaginary beamsplitters marked by $\mu$ and $\eta$.}\label{fig:scheme}
\end{figure}
The specific scheme we consider here is a Mach-Zehnder interferometer with double  detection (Fig.\,\ref{fig:scheme}), which can be viewed as a straightforward extension of the setup used in Ref.~\cite{20a1FrAgKhCh}. Here squeezed vacuum from an input degenerate optical parametric amplifier DOPA1 is injected into one of the input ports of a Mach-Zehnder interferometer. The second port is fed with the laser. The output signals of the interferometer, which carry the information about the phase shifts in the interferometer arms $\phi_1$, $\phi_2$, are amplified by two additional degenerate optical parametric amplifiers DOPA2, DOPA3 and measured by two  photodetectors. Finally, the two photocurrents are summed up with the optimal weight factors. In the current paper, we focus on the case of the direct detection, as shown in Fig.\,\ref{fig:scheme}. However, it can be shown that the double detection principle can be extended to the double homodyne detection case as well.

In the case of the direct detection, the sum of the phases $\phi_1+\phi_2$ is irrelevant. Therefore, we can assume without the generality loss that
\begin{equation}
  \phi_1 = -\phi_2 = \frac{\phi}{2} \,.
\end{equation}
We assume that our interferometer is a symmetric one, with the reflectivities/transmissivites matrices of both beamsplitters equal to
\begin{equation}
  \frac{1}{\sqrt{2}}\smatrix{1}{1}{1}{-1} .
\end{equation}
We take into account the internal (inside the interferometer) and external (at the interferometer ouputs, including the photodetectors inefficiency) loss by introducing imaginary beamsplitters with the power transmissivities, respectively, $\mu$ and $\eta$, which mix the probing light with the vacuum noise, see Fig.\,\ref{fig:scheme}.

In our calculations, we use the formalism of quadrature amplitudes, which are defined, for any annihilation operator $\hat{a}$, as follows \cite{Caves1985,Schumaker1985}:
\begin{equation}
  \hat{a}^c = \frac{\hat{a} + \hat{a}^\dagger}{\sqrt{2}} \,, \quad
  \hat{a}^s = \frac{\hat{a} - \hat{a}^\dagger}{i\sqrt{2}} \,,
\end{equation}
where the superscripts ``$c$'' and ``$s$'' denote the cosine and sine quadratures, respectively.

We decompose explicitly the incident laser field in the bright input port into the classical  amplitude $\alpha=\sqrt{N}$ and the noise part $\hat{z}_2$:
\begin{equation}
  \hat{a}_2 = \alpha+\hat{z}_2 \,,
\end{equation}
where $N$ is the mean number of quanta at this port. We assume without generality loss that $\alpha$ is real. In this case,
\begin{equation}
  \hat{a}_2^c = \sqrt{2}\alpha + \hat{z}_2^c \,, \quad \hat{a}_2^s = \hat{z}_2^s \,.
\end{equation}.

With an account for this, a straightforward calculation (see \eg \cite{17a1MaKhCh}) gives the input/output relations for the core part of the interferometer (from the first beamsplitter to the second one, including both of them):
\begin{subequations}\label{e}
  \begin{gather}
    \hat{e}_1^c = \delta\hat{e}_1^c \,, \quad
    \hat{e}_1^s = \mean{\hat{e}_1^s} + \delta\hat{e}_1^s \,, \\
    \hat{e}_2^c = \mean{\hat{e}_2^c} + \delta\hat{e}_2^c \,, \quad
    \hat{e}_2^s = \delta\hat{e}_2^s \,,
  \end{gather}
\end{subequations}
where
\begin{equation}
  \mean{\hat{e}_1^s} = \sqrt{2\mu}\,\alpha\sin\frac{\phi}{2} \,, \quad
  \mean{\hat{e}_2^c} = \sqrt{2\mu}\,\alpha\cos\frac{\phi}{2} \,
\end{equation}
are the classical components (the mean values) of the respective quadratures,
\begin{subequations}\label{io_quaf}
  \begin{gather}
    \delta\hat{e}_1^c = \sqrt{\mu}
      \biggl(\hat{a}_1^c\cos\frac{\phi}{2} - \hat{z}_2^s\sin\frac{\phi}{2}\biggr)
      + \sqrt{1-\mu}\,\hat{m}_+^c \,, \\
    \delta\hat{e}_1^s = \sqrt{\mu}
      \biggl(\hat{a}_1^s\cos\frac{\phi}{2} + \hat{z}_2^c\sin\frac{\phi}{2}\biggr)
      + \sqrt{1-\mu}\,\hat{m}_+^s \,, \label{e_1_s}\\
    \delta\hat{e}_2^c = \sqrt{\mu}
      \biggl(-\hat{a}_1^s\sin\frac{\phi}{2} + \hat{z}_2^c\cos\frac{\phi}{2}\biggr)
      + \sqrt{1-\mu}\,\hat{m}_-^c \,,\label{e_2_c} \\
    \delta\hat{e}_2^s = \sqrt{\mu}
      \biggl(\hat{a}_1^c\sin\frac{\phi}{2} + \hat{z}_2^s\cos\frac{\phi}{2}\biggr)
      + \sqrt{1-\mu}\,\hat{m}_-^s \,,
  \end{gather}
\end{subequations}
are the noise components with zero mean values,
\begin{equation}
  \hat{m}_\pm^{c,s} = \frac{\hat{m}_1^{c,s} \pm \hat{m}_2^{c,s}}{\sqrt{2}} \,,
\end{equation}
and $\hat{a}_{1,2}^{c,s}$, $\hat{e}_{1,2}^{c,s}$, and $\hat{m}_{1,2}^{c,s}$ are the quadrature operators of, respectively, the input modes, the output ones, and the vacuum noise modes associated with the internal losses, see notation in Fig.\,\ref{fig:scheme}.

It follows from Eqs.\,\eqref{e} that the corresponding photon-number operators are equal to
\begin{subequations}\label{n_e}
  \begin{gather}
    \hat{N}_{e1} = \frac{(\hat{e}_1^c)^2 + (\hat{e}_1^s)^2 - 1}{2}
      = \frac{\mean{\hat{e}_1^s}^2}{2} + \mean{\hat{e}_1^s}\delta\hat{e}_1^s
        + \mathcal{O}(\alpha^0) \,, \\
    \hat{N}_{e2} = \frac{(\hat{e}_2^c)^2 + (\hat{e}_2^s)^2 - 1}{2}
      = \frac{\mean{\hat{e}_2^c}^2}{2} + \mean{\hat{e}_2^c}\delta\hat{e}_2^c
        + \mathcal{O}(\alpha^0) \,,
  \end{gather}
\end{subequations}
where we denoted by $\mathcal{O}(\alpha^0)$ the small (second-order in quantum fluctuations $\hat{e}_{1,2}^{c,s}$) terms which do not contain $\alpha$. In the real-world high-precision interferometers, where the number of photons is large, $\alpha^2\gg1$, these terms can be neglected. In this approximation, the contributions of the quadratures $\hat{e}_1^c$ and $\hat{e}_2^s$ (which consist of the noise terms only and do not depend on $\alpha$) vanish, giving
\begin{subequations}\label{n_e_app}
  \begin{gather}
    \hat{N}_{e1}
      = \frac{\mean{\hat{e}_1^s}^2}{2} + \mean{\hat{e}_1^s}\delta\hat{e}_1^s \,, \\
    \hat{N}_{e2}
      = \frac{\mean{\hat{e}_2^c}^2}{2} + \mean{\hat{e}_2^c}\delta\hat{e}_2^c \,.
  \end{gather}
\end{subequations}
Two conclusions follow from this consideration. First, the quadratures $\hat{e}_1^s$ and $\hat{e}_2^c$, which remain in \eqref{n_e_app}, should be amplified by the output DOPAs to increase the signal. Second, the sine quadrature $\hat{a}_1^s$, which appears in Eqs.\,(\ref{e_1_s}, \ref{e_2_c}), should be squeezed by the input DOPA.

Taking these assumptions into account, we obtain the following equations for the quadratures $\hat{g}_1^s$, $\hat{g}_2^c$ of the effective fields at the photodetectors (with an account for the output losses and the detectors inefficiencies):
\begin{subequations}\label{g}
  \begin{gather}
    \hat{g}_1^s = \sqrt{\eta}\hat{e}_1^se^{r_2} + \sqrt{1-\eta}\,\hat{n}_1^s
      = \mean{\hat{g}_1^s} + \delta\hat{g}_1^s \,, \\
    \hat{g}_2^c = \sqrt{\eta}\hat{e}_2^ce^{r_2} + \sqrt{1-\eta}\,\hat{n}_2^c
      = \mean{\hat{g}_2^c} + \delta\hat{g}_2^c \,,
  \end{gather}
\end{subequations}
(note that it is these quadratures that contain the signal and appear in Eqs.\,\eqref{N1N2} below). Here,
\begin{equation}\label{mean_g}
  \mean{\hat{g}_1^s} = \sqrt{\eta}\mean{\hat{e}_1^s}e^{r_2} \,, \quad
  \mean{\hat{g}_2^c} = \sqrt{\eta}\mean{\hat{e}_2^c}e^{r_2}
\end{equation}
are the regular parts of $\hat{g}_1^s$, $\hat{g}_2^c$,
\begin{subequations}\label{delta_g}
  \begin{gather}
    \delta\hat{g}_1^s = \sqrt{\eta}\,\delta\hat{e}_1^se^{r_2} + \sqrt{1-\eta}\,\hat{n}_1^s
       \,, \\
    \delta\hat{g}_2^c = \sqrt{\eta}\,\delta\hat{e}_2^ce^{r_2} + \sqrt{1-\eta}\,\hat{n}_2^c
  \end{gather}
\end{subequations}
are the noise parts, and $\hat{n}_1^s$, $\hat{n}_2^c$ are the quadratures of the vacuum modes associated with the output losses.

\section{The phase sensitivity}\label{sec:optimization}

It follows from the equations (\ref{mean_g}, \ref{delta_g}) that the effective  (accounting for the detectors inefficiency) numbers of detected photons are equal to
\begin{equation}\label{N1N2}
  \hat{N}_1 = \frac{(\hat{g}_1^s)^2}{2} = \mean{\hat{N}_1} + \delta\hat{N}_1 \,,\quad
  \hat{N}_2 = \frac{(\hat{g}_2^c)^2}{2} = \mean{\hat{N}_2} + \delta\hat{N}_2 \,,
\end{equation}
where
\begin{equation}\label{mean_N}
  \mean{\hat{N}_1} = \frac{\mean{\hat{g}_1^s}^2}{2} \,, \quad
  \mean{\hat{N}_2} = \frac{\mean{\hat{g}_2^c}^2}{2} \,,
\end{equation}
and
\begin{equation}\label{delta_N}
  \delta\hat{N}_1 = \mean{\hat{g}_1^s}\delta\hat{g}_1^s \,, \quad
  \delta\hat{N}_2 = \mean{\hat{g}_2^c}\delta\hat{g}_2^c
\end{equation}
are the corresponding regular and  noise parts.

The explicit expressions for the mean numbers of quanta $\mean{\hat{N}_{1,2}}$, their variances $\mean{(\delta\hat{N}_{1,2})^2}$, the covariance $\mean{\delta\hat{N}_1\delta\hat{N}_2}$, as well as the corresponding values $\mean{\hat{N}_\pm}$, $\mean{(\delta\hat{N}_\pm)^2}$, $\mean{\delta\hat{N}_+\delta\hat{N}_-}$ for the sum and difference of $\hat{N}_{1,2}$:
\begin{equation}
  \hat{N}_\pm = \hat{N}_1 \pm \hat{N}_2 \,,
\end{equation}
are calculated in the Appendix \ref{app:N}. The following uncertainties of all relevant input quadratures are assumed:
\begin{subequations}\label{delta_xx}
  \begin{gather}
    \mean{(\hat{a}_1^s)^2} = \frac{e^{-2r_1}}{2}  \,, \\
    \mean{(\hat{z}_2^c)^2} = \frac{\mathcal{A}}{2} \,, \label{dz2c} \\
    \mean{(\hat{m}_+^s)^2} = \mean{(\hat{m}_-^c)^2}
      = \mean{(\hat{n}_1^s)^2} = \mean{(\hat{n}_2^c)^2} = \frac{1}{2} \,.
  \end{gather}
\end{subequations}
The factor
\begin{equation}
  \mathcal{A} = N(g^{(2)}-1) + 1 > 1
\end{equation}
in Eq.\,\eqref{dz2c}, where $g^{(2)}$ is the degree of second-order coherence, takes into account the contribution from the technical (super-Poissonian) noise of the laser light.

In order to calculate the phase measurment uncertainty, we will use the standard error-propagation formula:
\begin{equation}
  (\Delta\phi)^2
  = \frac{\mean{(\delta\hat{O})^2}}{\biggl(\partd{\mean{\hat{O}}}{\phi}\biggr)^2} \,,
\end{equation}
where $O$ is the measured quantity.

Consider now three different strategies of the phase measurement.

\paragraph{Single detector.}

The simplest approach is to use just one output, for example the first one:
\begin{equation}
  \hat{O} = \hat{N}_1 \,.
\end{equation}
In this case, Eqs.\,(\ref{mean_N_12}, \ref{delta_N_1}) give that
\begin{equation}\label{delta_phi_1}
  (\Delta\phi)^2 = (\Delta\phi_{\rm min})^2 + K\tan^2\frac{\phi}{2} \,,
\end{equation}
where
\begin{equation}\label{delta_phi_min}
  \Delta\phi_{\rm min} = \sqrt{\frac{e^{-2r_1}+\epsilon^2}{N}}
\end{equation}
is the best sensitivity achieved at $\phi=0$,
\begin{equation}\label{K}
  K = \frac{\mathcal{A}+\epsilon^2}{N}
\end{equation}
is the factor describing the sensitivity deterioration with the increase of $\phi$, and
\begin{equation}\label{eps2}
  \epsilon^2 = \frac{1-\mu}{\mu} + \frac{1-\eta}{\mu\eta}\,e^{-2r_2}
\end{equation}
is the overall quantum inefficiency of the interferometer, compare with Eqs.\,(5-7) of Ref.\,\cite{17a1MaKhCh}.

The factor $K$ defines the range of the phases where the sensitivity is close to $\Delta\phi_{\rm min}$. It follows from Eq.\,\eqref{delta_phi_1} that the full width at half minimum (FWHM) of this range is equal to
\begin{equation}
  \Delta_{\rm FWHM}
  = 4\arctan\sqrt{\frac{e^{-2r_1} + \epsilon^2}{\mathcal{A} + \epsilon^2}} \,.
\end{equation}
It is easy to see that the better is the sensitivity, the more narrow is the range where it can be achieved. In the most interesting high-sensitivity case of
\begin{equation}
  e^{-2r_1} \ll 1  \,, \quad \epsilon^2\ll 1 \hence
    \Delta\phi_{\rm min} \ll \Delta\phi_{\rm SNL} \,,
\end{equation}
this dependence can be expressed as follows:
\begin{equation}
  \Delta_{\rm FWHM} \approx \frac{4}{\sqrt{\mathcal{A}}}
    \frac{\Delta\phi_{\rm min}}{\Delta\phi_{\rm SNL}} \,.
\end{equation}

\paragraph{Differential detection.}

The second possible strategy is the measurement of the number of quanta difference:
\begin{equation}
  \hat{O} = \hat{N}_- \,.
\end{equation}
It could be justified by the fact that the sum number of quanta $\hat{N}_+$ does not depend on $\phi$, see Eq.\,\eqref{mean_N_pm}. In this case it follows from Eqs.\,(\ref{mean_N_pm}, \ref{delta_N_minus}) that
\begin{equation}\label{delta_phi_diff}
  (\Delta\phi)^2 = (\Delta\phi_{\rm min})^2 + K\cot^2\phi \,.
\end{equation}
Comparison of this result with Eq.\,\eqref{delta_phi_1} shows that this strategy is inferior to the simplest single-detection one, providing the same peak sensitivity, but twice as narrow FWHM range (albeit repeated twice in the same phases range $2\pi$):
\begin{equation}
  \Delta_{\rm FWHM}
   = 2\arctan\sqrt{\frac{e^{-2r_1} + \epsilon^2}{\mathcal{A} + \epsilon^2}} \,,
\end{equation}
or, in the case of $\Delta\phi_{\rm min}\ll1$,
\begin{equation}
  \Delta_{\rm FWHM} \approx \frac{2}{\sqrt{\mathcal{A}}}
    \frac{\Delta\phi_{\rm min}}{\Delta\phi_{\rm SNL}} \,.
\end{equation}

\paragraph{Optimal combination of two output sugnals.}

The best result can be obtained by combining both outputs with some optimized weight factor, which we denote as $k\pm1$:
\begin{equation}
  \hat{O} = \hat{N}_k = (k+1)\hat{N}_1 + (k-1)\hat{N}_2 = \hat{N}_- + k\hat{N_+}
  \,.
\end{equation}
Taking into account that $\hat{N}_+$ does not depend on $\phi$, the phase measurement error in this case is equal to
\begin{equation}
  (\Delta\phi)^2
  = \frac{\mean{(\delta\hat{N}_k})^2}{\biggl(\partd{\mean{\hat{N}_-}}{\phi}\biggr)^2} \,.
\end{equation}
Therefore, the value of $k$ that minimizes $\mean{(\delta\hat{N}_k)^2}$, provides also the best phase sensitivity. It follow from Eqs.\,(\ref{mean_N_pm}-\ref{delta_N_k}) that the the optimal value of $k$ and the corresponding value of $\Delta\phi$ are equal to
\begin{equation}\label{k_opt}
  k_{\rm opt}
  = -\frac{\mean{\delta\hat{N}_+\delta\hat{N}_-}}{\mean{(\delta N_+)^2}} = \cos\phi \,,
\end{equation}
and
\begin{equation}\label{delta_phi_opt}
  \Delta\phi = \Delta\phi_{\rm min}
\end{equation}
compare with Eqs.\,(\ref{delta_phi_1}, \ref{delta_phi_diff}). Thus, the optimal strategy \eqref{k_opt} gives a completely flat (independent on $\phi$) value of $\Delta\phi$.

At the same time, it creates the ``vicious circle'': the factor $k$ depends on $\phi$, which itself is the result of the measurement. Therefore, consider the suboptimal strategy, which uses the following value of $k$:
\begin{equation}\label{k_subopt}
  k_{\rm subopt} = \cos\phi_{\rm apr} \,,
\end{equation}
where $\phi_{apr}$ is the a priori mean value of $\phi$. It follows from Eqs.\,(\ref{mean_N_pm}, \ref{delta_N_k}) that in this case,
\begin{equation}\label{delta_phi_subopt}
  (\Delta\phi)^2 = (\Delta\phi_{\rm min})^2
    + K\frac{(\cos\phi - \cos\phi_{\rm apr})^2}{\sin^2\phi} \,.
\end{equation}

It follows from this equation that in order to obtain good phase sensitivity, $\Delta\phi\ll1$, the deviation of $\phi_{\rm apr}$ from the real $\phi$ also has to be small,
\begin{equation}\label{small_dphi_apr}
  \Delta\phi_{\rm apr} = \sqrt{\mean{(\phi_{\rm apr}-\phi)^2}} \ll 1 \,.
\end{equation}
However, actually this limitation is rather mild. Indeed, in the case of \eqref{small_dphi_apr}, Eq.\,\eqref{delta_phi_subopt} can be simplified as follows:
\begin{equation}\label{delta_phi_subopt1}
  (\Delta\phi)^2 = (\Delta\phi_{\rm min})^2 + K(\Delta\phi_{\rm apr})^2\,.
\end{equation}
Therefore, modest a priori knowledge of $\phi$, corresponding to
\begin{equation}\label{delta_phi_apr}
  \Delta\phi_{\rm apr}
    \lesssim \sqrt{\frac{e^{-2r_1} + \epsilon^2}{\mathcal{A} + \epsilon^2}} \,,
\end{equation}
allows one to obtain sensitivity close to the optimal one \eqref{delta_phi_opt} for all values of $\phi$. Note that the R.H.S. of the above equation does not contain $N$ and therefore could be much larger than, for example, $\Delta\phi_{\rm SNL}$.

\begin{figure}
  \includegraphics[scale=0.9]{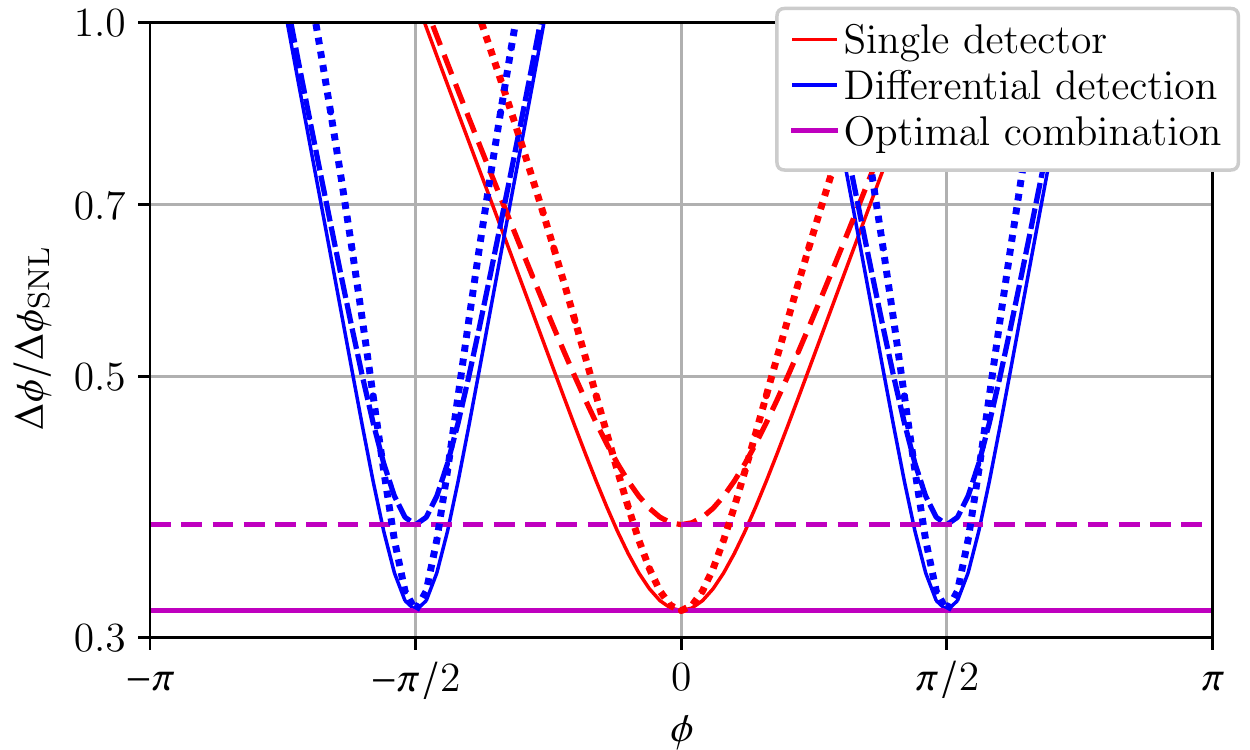}
  \caption{Normalized phase sensitivity for the three  considered measurement strategies. Solid lines: $\epsilon=0$, $\mathcal{A}=1$; dashed lines: $\epsilon=0.2$, $\mathcal{A}=1$; dotted lines: $\epsilon=0$, $\mathcal{A}=2$. In all cases, 10\,dB of the input squeezing is assumed ($e^{2r_1}=10$).}\label{fig:plots}
\end{figure}

In Fig\,\ref{fig:plots}, the phase measurement uncertainties for the considered strategies (\ref{delta_phi_1}, \ref{delta_phi_diff}, \ref{delta_phi_opt}) are plotted as functions of $\phi$.

\section{Conclusion}\label{sec:conclusion}

We analyzed here the sensitivity of a Mach-Zehnder interferometer with the squeezed light injected into one of the input ports and the parametric amplification of the both output signals, taking into account the optical losses and the detectors inefficiency. We compared three possible strategies of the measurement: (i) measurement of the number of photons at one of the interferometer output ports; (ii) measurement of the difference of photons numbers at the both outputs, and (iii) optimal combination of the numbers of photons measured at the two outputs. In all three cases, the output parametric amplifiers  suppress the effects of output losses and the photodetectors inefficiency. At the same time, while the first two strategies limit the range of phases where the high sensitivity could be achieved, the third one allows to beat the SNL by the same amount for all values of the phase $\phi$, assuming some modest a priori information on $\phi$. In particular, this information could be acquired using a preliminary low-precision [see Eq.\,\eqref{delta_phi_apr}] ``ranging'' measurement. In addition, the optimal combination of the two output signals allows to cancel the effect of the laser technical noise.

With a broadened range of high phase sensitivity, squeezing-assisted interferometers can improve the performance of many techniques where they have not been used so far. This is, first of all, Fourier-transform infrared (FTIR) spectroscopy, where the phase has to be scanned over ranges exceeding $2\pi$. Also, measurements of refractive index, used in various fields from aerodynamics to environmental gas sensing, are based on the phase shifts in a Mach-Zehnder interferometer and will benefit from broadening its high-sensitivity range.

\begin{acknowledgments}

{\sloppy

DKM and GS acknowledge financial support from DST, Govt.\,of India. DKM, GS, MC, and GF acknowledge financial support from DAAD, Germany under DST-DAAD joint project (DST/INT/DAAD/P-4/2019). DKM acknowledges financial support under EMR project (EMR/2016/001694) from SERB, New Delhi. FK acknowledges financial support from the Russian Science Foundation (project 20-12-00344).

}

\end{acknowledgments}

\appendix

\section{The photocounting statistics}\label{app:N}

It follows from Eqs.\,(\ref{mean_g}, \ref{delta_g}, \ref{delta_xx}), that
\begin{equation}
  \mean{g_1^s} = \sqrt{2}\,G\alpha\sin\frac{\phi}{2} \,, \quad
  \mean{g_2^c} = \sqrt{2}\,G\alpha\cos\frac{\phi}{2} \,,
\end{equation}
\begin{subequations}
  \begin{gather}
    \mean{(\delta\hat{g}_1^s)^2} = \frac{G^2}{2}\biggl(
        e^{-2r_1}\cos^2\frac{\phi}{2} + \mathcal{A}\sin^2\frac{\phi}{2} + \epsilon^2
      \biggr) , \\
    \mean{(\delta\hat{g}_2^c)^2} = \frac{G^2}{2}\biggl(
        e^{-2r_1}\sin^2\frac{\phi}{2} + \mathcal{A}\cos^2\frac{\phi}{2} + \epsilon^2
      \biggr) , \\
    \mean{\delta\hat{g}_1^s\,\delta\hat{g}_2^c}
      = \frac{G^2}{4}(-e^{-2r_1} + \mathcal{A})\sin\phi \,,
  \end{gather}
\end{subequations}
where
\begin{equation}
  G = \sqrt{\mu\eta}\,e^{r_1}
\end{equation}
is the amplitude transfer function of the setup and $\epsilon^2$ is given by \eqref{eps2}. Therefore, with account of Eqs.\,(\ref{mean_N}, \ref{delta_N}),
\begin{equation}\label{mean_N_12}
  \mean{\hat{N}_1} = G^2\alpha^2\sin^2\frac{\phi}{2} \,, \quad
  \mean{\hat{N}_2} = G^2\alpha^2\cos^2\frac{\phi}{2} \,,
\end{equation}
\begin{subequations}
  \begin{gather}
    \mean{(\delta\hat{N}_1)^2} = G^4\alpha^2\sin^2\frac{\phi}{2}\biggl(
        e^{-2r_1}\cos^2\frac{\phi}{2} + \mathcal{A}\sin^2\frac{\phi}{2} + \epsilon^2
      \biggr) , \label{delta_N_1} \\
    \mean{(\delta\hat{N}_2)^2} = G^4\alpha^2\cos^2\frac{\phi}{2}\biggl(
        e^{-2r_1}\sin^2\frac{\phi}{2} + \mathcal{A}\cos^2\frac{\phi}{2} + \epsilon^2
      \biggr) , \\
    \mean{\delta\hat{N}_1\,\delta\hat{N}_2}
      = \frac{G^4\alpha^2}{4}(-e^{-2r_1} + \mathcal{A})\sin^2\phi \,,
  \end{gather}
\end{subequations}
and
\begin{equation}\label{mean_N_pm}
  \mean{\hat{N}_+} = G^2\alpha^2 \,, \quad
  \mean{\hat{N}_-} = -G^2\alpha^2\cos\phi \,,
\end{equation}
\begin{subequations}\label{delta_N_pm}
  \begin{gather}
    \mean{(\delta N_+)^2} = \mean{(\delta\hat{N}_1)^2} + \mean{(\delta\hat{N}_2)^2}
      + 2\mean{\delta\hat{N}_1\,\delta\hat{N}_2}
      = G^4\alpha^2(\mathcal{A} + \epsilon^2) \,, \label{delta_N_plus} \\
    \mean{(\delta N_-)^2} = \mean{(\delta\hat{N}_1)^2} + \mean{(\delta\hat{N}_2)^2}
      - 2\mean{\delta\hat{N}_1\,\delta\hat{N}_2}
      = G^4\alpha^2(e^{-2r_1}\sin^2\phi + \mathcal{A}\cos^2\phi +\epsilon^2)
      \label{delta_N_minus} \,, \\
    \mean{\delta N_+\delta\hat{N}_-}
      = \mean{(\delta\hat{N}_1)^2} - \mean{(\delta\hat{N}_2)^2}
      = -G^4\alpha^2(\mathcal{A} + \epsilon^2)\cos\phi \,, \label{delta_N_pmpm}
  \end{gather}
\end{subequations}
\begin{multline}\label{delta_N_k}
  \mean{(\delta N_k)^2}
  = \mean{(\delta N_-)^2} + 2\mean{\delta N_+\delta\hat{N}_-}\cos\phi_{\rm apr}
    + \mean{(\delta N_+)^2}\cos^2\phi_{\rm apr} \\
  = G^4\alpha^2[
        (e^{-2r_1} + \epsilon^2)\sin^2\phi
        + (\mathcal{A} + \epsilon^2)(\cos\phi - \cos\phi_{\rm apr})^2
      ] \,.
\end{multline}



\end{document}